\documentclass[12pt,a4paper]{article}
\usepackage{amsmath,amssymb}
\usepackage{bm}
\usepackage{graphicx}
\usepackage[numbers,sort&compress]{natbib}

\begin{document}


\setlength{\baselineskip}{22pt}

\begin{flushleft}
\textbf{\Large{\textbf{Ultrafast electron dynamics at the Dirac node of the topological insulator Sb$_2$Te$_3$}}}
\end{flushleft}

\begin{flushleft}
Siyuan Zhu$^{1,*}$, Yukiaki Ishida$^{2}$, Kenta Kuroda$^{1}$, Kazuki Sumida$^{1}$,
Mao Ye$^{3}$, Jiajia Wang$^{4}$, Hong Pan$^{4}$, Masaki Taniguchi$^{1}$, Shan Qiao$^{3,4}$, Shik Shin$^{2}$ \& Akio Kimura$^{1}$
\end{flushleft}

\begin{flushleft}
\textit{$^1$Graduate School of Science, Hiroshima University, 1-3-1 Kagamiyama, Higashi-Hiroshima, Hiroshima 739-8526, Japan\\
$^2$Institute for Solid State Physics, the University of Tokyo, 5-1-5, Kashiwa-no-ha, Chiba 277-8581, Japan\\
$^3$State Key Laboratory of Functional Materials for Informatics, Shanghai, Institute of Microsystem and Information Technology, Chinese Academy of Sciences, 865 Chang Ning Road, Shanghai 200050, China\\
$^4$Department of Physics, State Key Laboratory of Surface Physics, and Laboratory of Advanced Materials, Fudan University, Shanghai 200433, China\\
}
\end{flushleft}

\noindent
\textbf{
Topological insulators (TIs) are a new quantum state of matter. Their surfaces and interfaces act as a topological boundary to generate massless Dirac fermions with spin-helical textures. Investigation of fermion dynamics near the Dirac point is crucial for the future development of spintronic devices incorporating topological insulators.
However, research so far has been unsatisfactory  because of a substantial overlap with the bulk valence band  and a lack of a completely unoccupied Dirac point (DP).
Here, we explore the surface Dirac fermion dynamics in the TI Sb$_2$Te$_3$ by time- and angle-resolved photoemission spectroscopy (TrARPES).
Sb$_2$Te$_3$ has a DP located completely above the Fermi energy ($E_F$) with an in-gap DP.
The excited electrons in the upper Dirac cone stay longer  than those below the Dirac point to form an inverted population.
This was attributed to a reduced density of states (DOS) near the DP . 
}

\vspace{10mm}
\newpage
Three-dimensional TIs have emerged as a new state of condensed matter and are characterized by nontrivial gapless surface states (SS) that occur because of a strong spin--orbit coupling.
The SS traversing the band gap between the bulk valence band (VB) and conduction band (CB) can be described by the Dirac equation for massless fermions~\cite{FuKaneMele, HZhang_SCZhang, YLChen, XiaHasan, SCZhang_Rev}.
Additionally, the SS are spin-polarized and the spin orientations are fixed with respect to their momenta~\cite{Hsieh_Spin1, Hsieh_Spin2, Roushan_Nature}. Such a peculiar electronic structure, which originates from its $\pi$ Berry phase, results in an anti-localization of surface electrons with a suppressed backscattering probability. A number of 3D TIs, including Bi$_2$Se$_3$, Bi$_2$Te$_3$, Sb$_2$Te$_3$, TlBiSe$_2$, PbBi$_2$Te$_4$ and SnSb$_2$Te$_4$, has been discovered experimentally ~\cite{Xia_NP, Hsieh_PRL1, Sato_PRL, Kuroda_PRL, Niesner_PRB}. TIs have recently attracted much attention because of their possible applications in spintronic devices and in ultra-fast and fault tolerant quantum computation ~\cite{Wolf_Science, Kong_Nanotech, Xiu_Nanotech,Mclver_Nanotech}. When aiming to improve such novel device applications incorporating TIs, it is important to understand the hot carrier dynamics of the surface Dirac fermions.

Angle resolved photoemission spectroscopy (ARPES) implemented by a pump-and-probe method is a powerful tool to study the unoccupied states and electron dynamics with energy and momentum resolutions.
Many groups have made great processes of TrARPES on TIs~\cite{Johannsen_TrPES, Sobota_TrPES, Sobota_TrPES2, Kirilyuk_RMP, Hsieh_PRL2, Wang_PRL, Crepaldi_PRB, Wang_Science, Sobota_TrPES3, Hajlaoui_NatCom, Luo}.
Recently, TrARPES measurements at the sub-20-meV energy resolutions became possible~\cite{Kim_TrPES, Ishida_TrARPES}. This enabled us to observe the electron dynamics near the DP in detail. 
To examine the flow of electrons across the DP, we need an initial state situation (for example, before pumping) in which both the upper and lower parts of the Dirac cone are empty.
This could be realized in $p$-type TIs, wherein the DP is located above $E_F$. Graphene, whose DP is almost at or below $E_F$, is therefore not suitable for this purpose.
The $p$-type Bi$_2$Se$_3$ is also unsuitable because the lower part of Dirac cone is not energetically isolated from the bulk valence band~\cite{Sobota_PRL}.
This feature can also be seen from the absence of the Landau level quantization  in the lower part of the surface Dirac cone, while it is visible above the DP~\cite{Cheng_PRL, Hanaguri_PRB}.

In contrast, Sb$_2$Te$_3$\ shows surface Landau quantizations over the energy range of $\sim$240~meV (120~meV below and 120~meV above the DP)~\cite{Jiang_PRL, Jiang_PRL2}. Here, the Dirac cone SS is separated from the bulk states, which enables us to study an isolated Dirac cone. Second, a Sb$_2$Te$_3$ single crystal is naturally $p$-doped, and the DP is located above the $E_F$. Therefore, we do not need to dope any element into the mother crystal. This is advantageous when attempting to increase the quality of the sample. Having considered the above-mentioned characteristics, Sb$_{2}$Te$_{3}$ is suitable for the present study.

In this study, we investigated the unoccupied bulk and surface states of Sb$_2$Te$_3$ using TrARPES. The electron dynamics below and above the DP were also revealed. One of the most striking findings is that the decay of the pump-induced carriers are bottlenecked at the DP, so that the hot carriers in the upper part of the SS stay longer  than those in the lower part.

\vspace{10mm}
\noindent
\textbf{Results and discussion}

By pumping the electrons into the unoccupied side, we observed a linear Dirac cone SS as shown in Fig.~1(a). Here, the pump-and-probe delay, $t$, was set to 0.4 ps. The DP is located $\sim$180~meV above the $E_{\rm F}$ and the Dirac velocity was estimated to be $\sim$$2.3\times10^5$ m/s. We found that both the upper and lower parts of the Dirac cone (UDC and LDC, respectively) were clearly visible above $E_F$ and they do not overlap with the bulk continuum states. Figure~1(b) shows the constant energy contours at 100, 290 and 410~meV. With increasing the energy, the SS evolves from a circular to hexagonal shape. The isotropic constant surface can be observed both below and above the DP within the bulk energy gap. The hexagonal warping of the constant energy surfaces is quite small as long as bulk continuum states do not overlap with the SS. In the previous STM  study on Sb$_2$Te$_3$, the DP is 80~meV higher, whereas the energy range of the SS ($\sim$120~meV above and below the DP) is consistent with the present observations~\cite{Jiang_PRL, Jiang_PRL2}; see Fig.~1(a). With such an ideal situation, there is a good opportunity to study the carrier dynamics of UDC and LDC separately, where interference from the bulk states is minimized.

To study the pump-induced dynamics of the surface Dirac fermions, we altered the pump and probe delay and investigated the time dependent variations in the TrARPES images.  Figure~2(a) shows the difference image along the $\overline{\Gamma}-\overline{K}$ line  measured at $t$ = 0.4~ps. Both the Dirac cone SS and unoccupied bulk state were clearly observed. To show the energy dependent dynamics, we set energy and momentum frames [A to I: see Fig.~2(a)] and plotted the intensity variation in each frame as functions of $t$ [see Fig.~2(b)]. Also, to show the variation in the different bands more clearly, we show the original and difference images for typical delay times in Fig.~2(c) and in a supplementary movie S4. 
Here we note that the intensity variation line profiles of bulk and surface states at the same energy overlapped each other as shown in the supplementary Fig.~S1.  

In the highest energy region A, we observed a fast rise of intensity that was limited by the time resolution without significant delay. The intensity variation was almost symmetric about $t$ = 0. This indicates that the intensity variation in region A comprises instantaneous filling of the states by direct excitations and very fast flow of the excited electrons out of region A into the lower energy states. Because the flow of electrons into region A from higher energies is negligibly small, the line shape does not show significant asymmetric tailing into $t > 0$.

Next, we investigated the energy regions A,  B, C, D, E and F, which are overlapped to the conduction band. The duration of the intensity variation became long as the DP was approached. This indicates that there was an energy dependence on the transfer rate of electrons: The net flow rate of electrons from high to low energy decreased when the Dirac point was approached. This can occur because the available phase space diminishes on the approach of the DP, and so the hot carriers pile up around the bottom of the UDC. Similar behaviour was observed above the DP for Bi$_2$Se$_3$~\cite{Sobota_TrPES}. Considering that the behaviour can be represented by an exponential decay, the decay constant, $\tau$, of the different regions varied from 0.2 to 2~ps, which is comparable to a recent study on Sb$_2$Te$_3$~\cite{Reimann_PRB}.

The most striking observation was found across the DP, namely in the intensity variations of regions G and H.
Although region H in the LDC was located lower in energy than G in the UDC, the intensity after $\sim$1~ps diminished faster in H than in G as shown in 
Fig.~2(b).
Figure~2(d) shows EDCs (integral of the TrARPES images over $\pm$15~degrees) normalized to the peak in the LDC region. 
From $\sim$0.4  to $\sim$3~ps, the spectral intensity in the UDC region is higher than that in the LDC region. We take this as evidence for the population inversion across the DP. Note, if the electron distribution was obeying a thermal Fermi-Dirac function, there would be no crossings between the intensity variation line profiles at different energies, which is opposed to the case seen in Fig.~2(b); also see supplementary Fig.~S2. After $\sim$3~ps, the intensity in the UDC region becomes smaller than that in the 
LDC region [right panel of Fig.~2(d)]. Correspondingly, the intensity variation line profiles of regions G and H shown in Fig.~2(d) almost overlap each other after $\sim$3~ps.  

The population inversion can occur across the DP because the node acts as a bottleneck for the electrons flowing from high to low energies: The low DOS near the DP is considered to play a key role in the formation of the population inversion. In order to support this view, we solved a rate equation under DOS having some structures. We find that an inverted population can be formed when the DOS has a valley-like structure similar to the case having a DP; see Fig.~S3 in the supplementary file. The simulation also shows that, after the `electron jam' near the node is cleared, the decay profiles across the node become similar, which qualitatively explicates the decay-profile behaviour after $\sim$3~ps seen in Fig.~2(b). 

We also observed that the rise time of the intensity in region I, which is close to $E_F$, is faster than those in the UDC regions. The fast intensity rise around $E_F$ is attributed to the impact ionization: The direct photo-excitation accompanies the low energy excitations across $E_F$~\cite{Sze,Levinshtein,Nazarov}. The effect of impact ionization is limited to $\lesssim$50~meV and is similar to the Fermi cutoff  broadening, as seen in time-resolved photoemission spectra of metals~\cite{Au}. Because the effect of impact ionization occurs only in the vicinity of $E_F$, the carrier dynamics in the SS are less affected by the impact ionization.

Schematics of the pump and decay processes from the state before pumping to the final state are shown in Fig.~3. As shown in Fig.~3(b), the direct photo-excitation from the occupied to the unoccupied states is accompanied by the impact ionization. During the decay [Fig.~3(c)], the flow of electrons from high to low energy is bottlenecked near the DP to result in the hourglass-shaped electron distribution shown in Fig.~3(d).

\vspace{10mm}
\noindent
\textbf{Conclusion}

The conclusion is three-fold. First, TrARPES on Sb$_{2}$Te$_{3}$ revealed the surface state Dirac cone in the unoccupied region. It was found to be isotropic within the bulk energy gap and the Dirac velocity was larger than that of Bi$_{2}$Se$_{3}$. Second, a rapid intensity increase was found near $E_F$, which was caused by the creation of a large number of low energy electron-hole pairs due to impact ionization. Third, we found the spectral intensity inversion at $\sim$0.4 to $\sim$3~ps across the DP. The population inversion across the Dirac dispersion may be used as an optical gain medium for broad band lasing if the duration of the inversion can be elongated \cite{GrapheneLaser}, for example, by continuously injecting carriers into the UDC. 

\vspace{10mm}
\noindent
\textbf{Methods:}

The Sb$_2$Te$_3$ single crystal was grown by the Bridgeman method. The results of electron probe micro analysis (EPMA) showed an atomic ratio of Sb:Te $=2.03:2.97$. The experiment was performed with linearly polarized 5.98  (probe) and 1.5~eV (pump) pulses derived from a Ti:sapphire laser system operating at a repetition rate of 250~kHz~\cite{Ishida_TrARPES}. The photoelectron kinetic energy and emission angle were resolved using a hemispherical electron analyser . The measurement was done at 8~K with an energy resolution of $\sim$15~meV. The origin of the pump-and-probe delay ($t$ = 0) and the time resolution of 250~fs was determined from the TrARPES signal of graphite attached next to the sample. The spot diameters of the pump and probe were 0.5 and 0.3~mm, respectively.

\noindent
\textbf{Acknowledgements:}

\noindent
The TrARPES measurements were jointly carried out by the Laser and Synchrotron Research Center of the Institute for Solid State Physics and the University of Tokyo (Proposal No. A181, A184). This work was partly supported by KAKENHI (26247064 and 26800165) and the Grant-in-Aid for Scientific Research (A) of the Japan Society for the Promotion of Science.



The authors declare no competing financial interests.

\newpage
\noindent
\textbf{Figure Legends}
\setlength{\baselineskip}{22pt}

\vspace{10mm}
\noindent
\textbf{\label{fig_1}Figure 1: Band structure of Sb$_2$Te$_3$ revealed into the unoccupied side.}
(a) The TrARPES images of Sb$_2$Te$_3$ recorded at $t$ = 0.4~ps along the $\overline{\Gamma}-\overline{K}$ direction. (b) Constant energy maps at 100, 290 and 410~meV. 

\vspace{10mm}
\noindent
\textbf{\label{fig_2_}Figure 2: TrARPES of Sb$_2$Te$_3$.}
(a) TrARPES images recorded along the $\overline{\Gamma}-\overline{K}$ 
line recorded before pump (left; images recorded at $\le$0.6~ps were averaged), at 0~ps (middle), and their difference (right panel). The frames A to I span in the angular range of $\pm$15 degrees and in the energy ranges of [0.80, 0.90], [0.70, 0.76], [0.60, 0.66], [0.50, 0.56], [0.40, 0.46], [0.30, 0.36], [0.20, 0.26], [0.06, 0.17] and [0.01, 0.05] (in units of eV), respectively. (b) Intensity variation line profiles. Integrated intensity in each of the frames A to I is plotted as functions of delay time in a linear (upper panel) and in a logarithmic scale (lower panel). 
(c) TrARPES images. Upper and lower panels show TrARPES and difference to that recorded before pump. Full set of TrARPES and difference images are provided as a supplementary movie S4. (d) EDCs (integration of TrARPES images over $\pm$15 degrees) recorded at 0 $\le t \le$ 1.00~ps (left), 1.00 $\le t \le$ 2.97~ps (middle), and at 2.97 $\le t \le$ 5.13~ps (right panel). Here, the EDCs are normalized to the area around the peak in the LDC region. For the full set of EDCs, see supplementary movie S5. 

\vspace{10mm}
\noindent
\textbf{\label{fig_3}Figure 3: Schematics of the pump and decay processes.}
The state before pumping (a), upon the pump (b), subsequent decay (c) leading to an hourglass-shaped electron distribution (d), and the final state (e). The colour gradation represents the electron density.

\newpage
\begin{figure*}
	\centering
	\includegraphics[width=8.7cm]{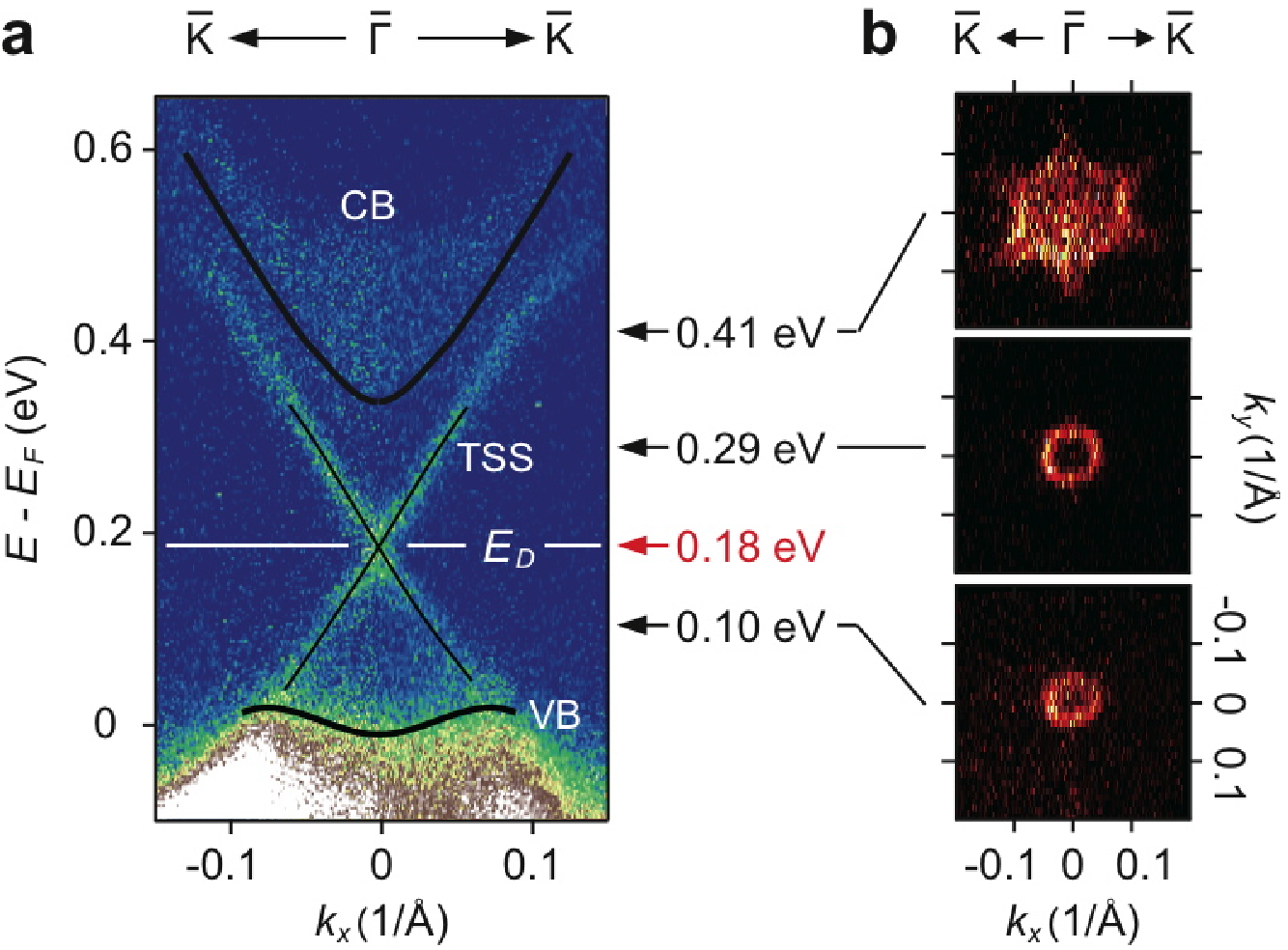}
    \caption{\label{fig:1}}
\end{figure*}

\newpage
\begin{figure*}
	\centering
	\includegraphics[width=13cm]{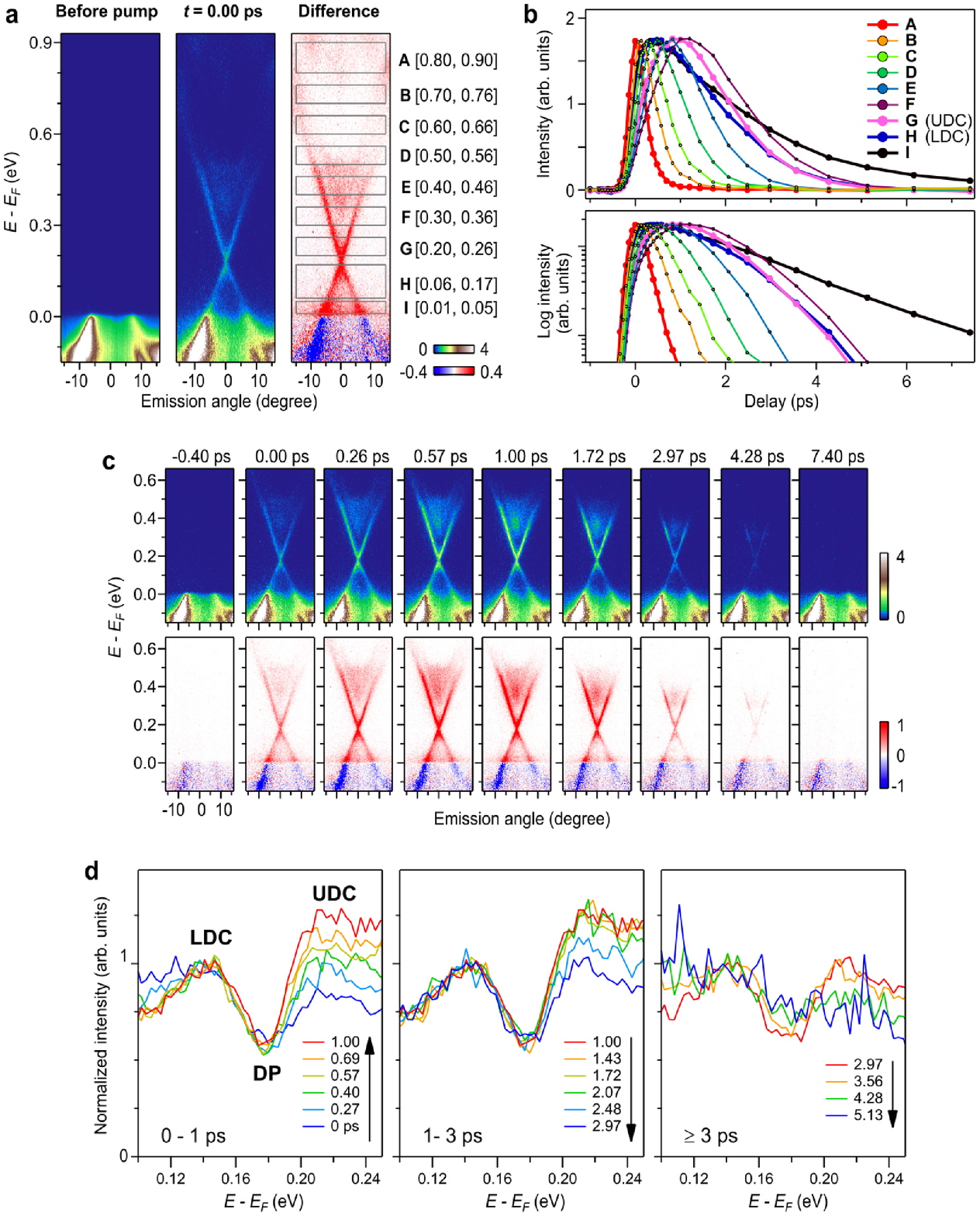}
    \caption{\label{fig:2}}
\end{figure*}

\newpage
\begin{figure*}
	\centering
	\includegraphics[width=13cm]{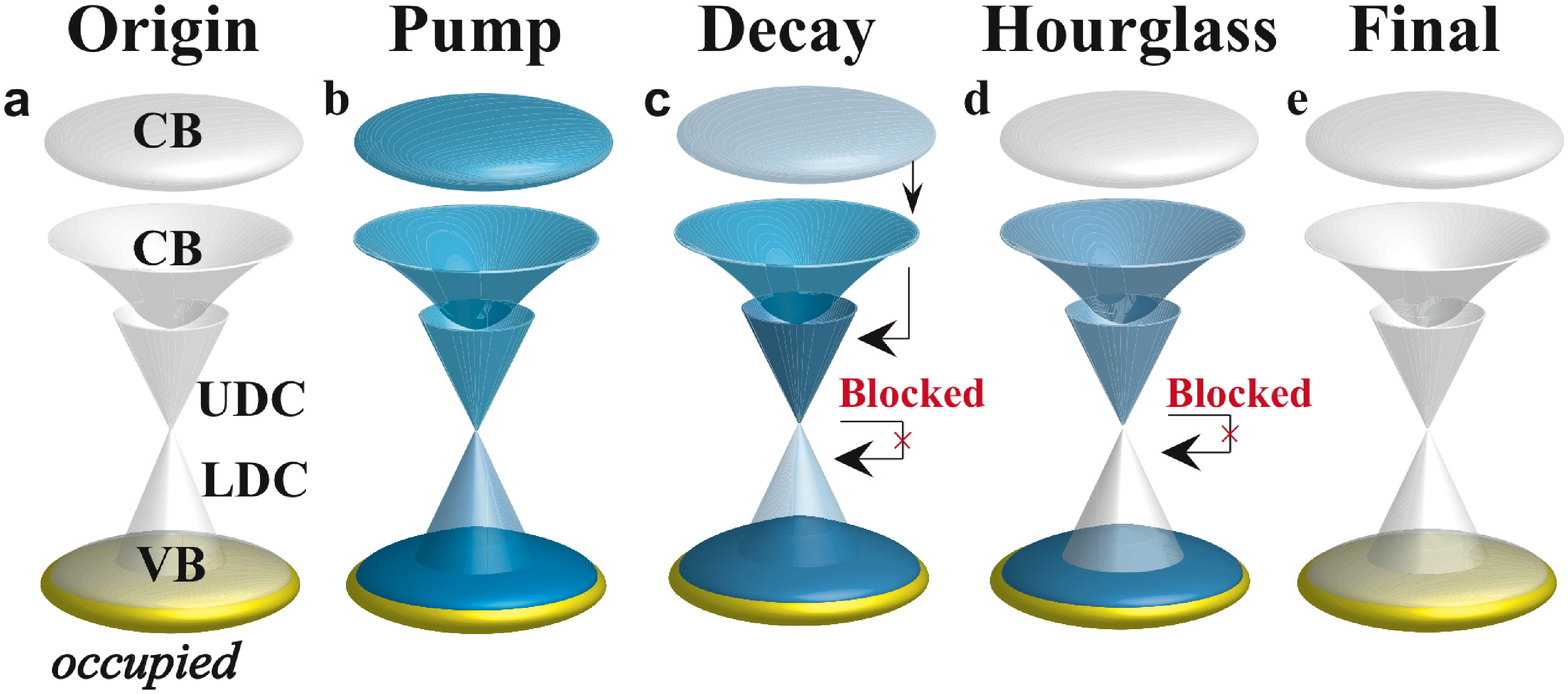}
	\caption{\label{fig:3}}
\end{figure*}

\end{document}